# New Common Proper-Motion Pairs from the PPMX Catalog

Rafael Caballero, Blanca Collado-Iglesias, Sara Pozuelo-González,
Antonio Fernández-Sánchez

Agrupación Astronómica Hubble,
Martos, Jaén, Spain

Email: rafa@sip.ucm.es

**Abstract:**  We use data mining techniques for finding 82 previously unreported common proper motion pairs from the PPM-Extended catalogue. Special-purpose software automating the different phases of the process has been developed. The software simplifies the detection of the new pairs by integrating a set of basic operations over catalogues. The operations can be combined by the user in scripts representing different filtering criteria. This procedure facilitates testing the software and employing the same scripts for different projects.

## Introduction

In previous papers (Caballero 2009, Caballero 2010) we data mined different catalogs using some criteria to obtain new common proper-motion pairs (CPMP's from now on) not included in the WDS (Washington Double Star Catalog, Mason, et al., 2003). This idea is not new and has been used for instance by Greaves (2004).

During the development of the projects it became clear that the process was almost the same in all the cases, with a few changes due to the particular characteristics of each catalog. Therefore it seemed interesting to develop a special-purpose software. Such a project was suggested as a Master's thesis topic at the faculty of Computer Science at the University Complutense of Madrid (Spain). The project was developed by Blanca Collado-Iglesias, Sara Pozuelo-González and Antonio Fernández-Sánchez, and directed by Rafael Caballero. This paper presents 82 new CPMPs from the PPM-Extended catalog (PPMX, see Röser, 2008) obtained with the help of this application.

## The Data Mining Process

The overall data mining process can be described as follows:

1.  Downloading (part of) the main catalog $C$ , usually from the online VizieR Service web page (Allende & Dambert 1999). Sometimes portions of auxiliary catalogs are also needed, for instance to complete the information about spectra, visual magnitude, etc.

2.  Importing $C$ data into a relational database such as Access or MySQL (and also the auxiliary catalogs).

3.  Obtain the Cartesian product $D = C \times C$.   $D$ is thus a table of pairs.

4.  Delete from D all the pairs with separation greater than some arbitrary number, for instance 100 seconds.

5.  Remove from D all the pairs that are already part of the WDS.

6.  Apply some criteria, in order to keep in D only possible CPMPs. A typical case is the Halbwachs' criteria (Halbwachs, 1986).

7.  If possible, introduce further criteria that can help to increase the data quality, i.e. to reject those pairs that are more likely not physically attached.

8.  Finally, check every pair in the photographic plates available at ALADIN (Bonnarel et al., 2000), looking for two stars with noticeable motion and roughly the same astrometry data in the expected position.

## New Common Proper-Motion Pairs from the PPMX Catalog

9.   Complete the data with astrometry and other suitable data from auxiliary catalogs.

## The Software Application

The opening screen of the application is shown in Figure 1.  It was developed in the Java programming language. This language was chosen because it easily allows connecting to different databases using a convenient JDBC (Java Database Connectivity) driver. Initially, the system is configured for using the relational database MySQL, but it can be readily adapted for other databases such as Oracle or Access. The program allows the user importing catalogs obtained from Vizier in text format with the fields separated by ";". By parsing the header produced by VizieR, the type and size of the different attributes is detected, and a suitable SQL table created.

Other options for general management of catalogs are included. Probably one of the most useful ones is the "Join Catalogs" option. With this option the user can cross two tables containing individual stars yielding a new table containing those pairs with separation less than a parameter in seconds (steps 3 and 4 of the overall process described in the previous section). This table will contain the set of initial candidates.

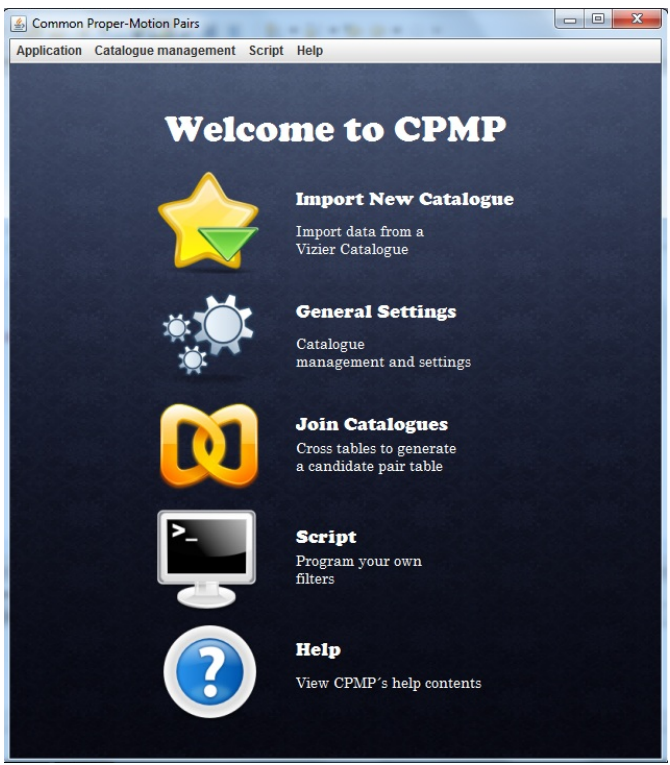

**Figure 1**: Application main screen.

However, the most appealing feature of the program is that it allows the users to program their own scripts for filtering the candidate pairs. The scripts can be saved and are defined by combining a few core operations, like adding a new field to an existing table, setting the value for some field for all the rows fulfilling some condition, combining catalogs to produce a new one, or deleting all the rows that do not satisfy a given condition. Furthermore, the system allows defining parameterized functions that will be used from different scripts. As a simple example consider the following user-defined function:

> *function* setFilterPM(2)
> *begin*
>   temp1<-*newAttribute*($1,mu,double);
>   temp2<-*newAttribute*(temp1,b_mu,double);
>   temp3<-*attribute*(temp2,mu,'sqrt
> (pmra*pmra+pmde*pmde)')[true];
>   temp4<-*attribute*(temp3,b_mu,'sqrt
> (b_pmra*b_pmra+b_pmde*b_pmde)')[true];
>    $2<-*filter*(temp4)[mu>=50 and b_mu>=50];
> *end*

The function receives as input a table $1 of pairs, which is assumed to contain fields *pmra*, *pmde*, *b_pmra*, and *b_pmde*, which represent the proper motion in RA and DEC of the two components, and it creates a new table $2 such that:

- It contains two new fields *mu* and *b_mu* such that for each row $mu = \sqrt{(pmra^2 + pmde^2)}$ and $b\_mu = \sqrt{(b\_pmra^2 + b\_pmde^2)}$.
- It only contains those rows verifying $mu>=50$ and  $b\_mu>=50$ (i.e. both components with proper motion over 50 millisecond of arc/year).

The function is self-explanatory: the two first statements after the reserved word *begin* add the new fields, both of type *double* (real numbers). The two next lines give value to the new fields. The condition *[true]* at the end of each statement specifies that the new values must affect all the rows. Finally, the last statement before the reserved word *end* removes all the rows corresponding to pairs where some of the components have proper motion below 50 mas/yr.

## A Test Case: the PPMX Catalog

In order to check the software the PPMX catalog was chosen.  In particular we started downloading the catalog for entries with available V magnitude and with proper motion over 50 milliseconds of arc per

## New Common Proper-Motion Pairs from the PPMX Catalog

year. Then we used the option "join catalogs" to produce the initial set of candidate pairs. 2171 pairs with separation under 100 seconds were obtained. Then a new, more restrictive, version of the Halbwachs' criteria was applied (see next section), further reducing the set of pairs to 979. Then the pairs already in the WDS, or those likely to be included in the catalog in the near future, were removed leaving 85 pairs. Another three pairs were excluded after using the Reduced Proper Motion (RPM) discriminator proposed by Salim & Gould (2003). Finally, all the pairs were checked in the photographic plates, finding all of them. The summary of this process is shown in Table 1.

**Table 1**: Summary of the data mining process for PPMX

| Phase | Rows |
|---|---|
| PPMX initial subset | 172115 |
| Candidate pairs | 2171 |
| After (modified) Halbwachs criteria | 979 |
| After removing pairs already in WDS | 213 |
| After removing pairs in other lists | 85 |
| After RPM criterion | 82 |
| After checking photographic plates | 82 |

## Halbwachs revisited

The three criteria originally proposed by Halbwachs for distinguishing physical and optical pairs from their proper motion are:

1. $(\mu_1 - \mu_2)^2 < -2\,(\sigma_1^2 + \sigma_2^2)\,\ln(0.05)$
2. $|\mu_1|,\ |\mu_2| \geq 50$ mas/yr
3. $\rho\,/\,|\mu_1|,\ \rho\,/\,|\mu_2| < 1000$ yr

where $\mu_1$, $\mu_2$ are the two proper motion vectors, $\sigma_i$ is the mean error of the projections on the coordinate axes of $\mu_i$, and $\rho$ is the angular separation of the two stars. The first condition checks if the hypothesis $\mu_1 = \mu_2$ is admissible with a 95% confidence considering the given errors $\sigma_1$ and $\sigma_2$. If $\mu_1 = (\mu_{11}, \mu_{12})$, $\mu_2 = (\mu_{21}, \mu_{22})$ then this condition can be rewritten as

(1') $(\mu_{11} - \mu_{21})^2 + (\mu_{21} - \mu_{21})^2 < -2\,(\sigma_{12} + \sigma_{22})\,\ln(0.05)$

However in previous experiences it was observed that this criterion allowed pairs with noticeably different values in some axes, and thus some additional criterion was needed. In this project we propose using the condition for each axis separately, i.e. to replace condition (1) (or (1')) by:

(1.a) $(\mu_{11} - \mu_{21})^2\ \ < -2\,\sigma_{12}\,\ln(0.05)$
(1.b) $(\mu_{12} - \mu_{22})^2\ \ < -2\,\sigma_{22}\,\ln(0.05)$

It is straightforward to check that the conjunction of (1.a) and (1.b) imply (1'), and therefore the new condition is more restrictive. In particular, in the case of the PPMX project replacing the condition (1) by (1.a) and (1.b), results in 15 additional pairs filtered out. These pairs are precisely those with noticeable differences in any axis. Hence, we think that substituting (1) by (1.a) and (1.b) is a good practice that improves the quality of the results.

## Results

Table 2 shows the final list with the new CPMP obtained. The astrometry (precise coordinates, position angle, separation, and date) have been obtained from the 2MASS catalog. The visual magnitudes correspond to the values in PPMX. The pair separation ranges from 8.64" to 99.40", and the visual magnitudes are between 6.25 and 13.07. The pair at RA-dec 21 36 58.08 -35 53 02.93 is especially remarkable, since the parallax of both components is known and compatible: 14.95 ±0.93 mas for A and 15.09 ±1.21 mas for B. In this case it seems quite safe to say that this bright pair (mags. 7.35 and 8.60) is physically attached. For another 15 pairs the parallax of the primary is known, and for a larger number the spectral type of one or the two components is available either from Hipparcos (Perryman, 1997) or from Tycho-2 (Wright, 2003). In three cases the secondary of the new pair is a close pair already in WDS, see notes 13, 21, and 34.

Table 3 contains the proper motion data of the new CPMP. The data corresponds to PPMX.

## Conclusions

Ensuring the data quality must be one of the main goals of any data mining project. With this purpose we have developed a software application that simplifies the different phases of the project. The filters for selecting and reducing the number of candidate pairs are easily prepared by the user by employing a small subset of basic operations over catalogues, which constitutes the core language offered by the application. The

## New Common Proper-Motion Pairs from the PPMX Catalog

| Discovery Designation | RA DEC | | Mags | | PA | SEP | DATE | NOTES |
|---|---|---|---|---|---|---|---|---|
| CBL | 00 32 19.099 | −21 50 33.69 | 11.7 | 12.29 | 208.45 | 49.78 | 2000.783 | |
| CBL | 00 55 33.213 | −43 16 11.73 | 12.24 | 12.63 | 134.53 | 28.20 | 1999.713 | |
| CBL | 01 00 52.512 | −18 56 57.08 | 11.44 | 11.57 | 220.88 | 25.25 | 1998.626 | |
| CBL | 01 52 27.571 | −49 31 25.56 | 9.04 | 9.16 | 207.56 | 42.41 | 1999.812 | (1) |
| CBL | 02 04 18.761 | −70 59 40.88 | 10.48 | 11.77 | 103.88 | 24.05 | 1999.894 | |
| CBL | 02 05 27.334 | +38 20 57.02 | 12.29 | 13.07 | 130.16 | 31.88 | 1998.810 | |
| CBL | 03 10 41.550 | −20 06 41.54 | 7.62 | 10.61 | 313.69 | 59.79 | 1998.878 | (2) |
| CBL | 03 24 54.681 | −43 12 55.77 | 9.86 | 11.62 | 192.09 | 26.63 | 1999.648 | (3) |
| CBL | 03 46 09.569 | −41 12 22.33 | 9.25 | 11.48 | 257.08 | 66.28 | 1999.629 | (4) |
| CBL | 04 23 48.192 | −76 43 09.45 | 11.13 | 11.29 | 265.20 | 45.21 | 1998.840 | |
| CBL | 05 01 36.173 | −44 49 49.05 | 7.55 | 10.58 | 265.53 | 50.77 | 1999.722 | (5) |
| CBL | 05 23 39.978 | −38 18 48.09 | 11.82 | 12.20 | 193.43 | 24.06 | 1999.173 | |
| CBL | 05 57 24.758 | −40 23 51.78 | 8.15 | 11.28 | 350.10 | 45.10 | 1999.190 | (6) |
| CBL | 06 15 01.664 | −63 20 38.82 | 11.71 | 12.57 | 219.59 | 15.59 | 1998.947 | |
| CBL | 06 38 17.669 | +18 28 24.58 | 11.44 | 11.79 | 63.63 | 55.81 | 1997.898 | |
| CBL | 07 03 08.446 | −73 50 13.91 | 11.12 | 11.34 | 140.15 | 51.35 | 2000.167 | |
| CBL | 07 08 55.244 | −11 23 29.58 | 11.22 | 12.32 | 31.21 | 48.15 | 1999.138 | |
| CBL | 09 33 19.911 | −07 11 24.75 | 6.25 | 10.78 | 19.14 | 57.81 | 1999.048 | (7) |
| CBL | 09 51 08.004 | −18 39 31.43 | 7.37 | 10.83 | 115.95 | 50.68 | 1998.322 | (8) |
| CBL | 10 14 38.104 | −13 33 29.10 | 9.04 | 10.34 | 177.99 | 28.62 | 1999.299 | (9) |
| CBL | 10 15 08.028 | −65 26 11.04 | 11.15 | 11.47 | 304.57 | 31.11 | 2000.230 | |
| CBL | 10 16 15.352 | −17 11 13.12 | 10.67 | 11.78 | 282.61 | 41.57 | 1998.234 | |
| CBL | 10 32 03.297 | −30 28 05.49 | 11.46 | 11.78 | 29.56 | 14.49 | 1999.223 | |
| CBL | 10 47 27.741 | −11 54 08.72 | 9.83 | 10.46 | 48.64 | 27.76 | 1998.256 | (10) |
| CBL | 11 29 03.259 | −38 17 05.77 | 9.48 | 11.35 | 109.61 | 38.72 | 1999.272 | (11) |
| CBL | 11 35 52.047 | −40 40 36.43 | 11.85 | 12.18 | 247.31 | 20.08 | 1999.275 | |
| CBL | 11 53 22.088 | −67 07 05.56 | 10.78 | 11.88 | 119.24 | 36.11 | 2001.121 | |
| CBL | 12 05 25.156 | +17 17 21.69 | 7.63 | 10.04 | 311.55 | 43.20 | 1998.033 | (12) |
| CBL | 12 13 30.463 | −48 47 46.64 | 8.80 | 10.57 | 347.83 | 49.99 | 1999.357 | (13),(14) |
| CBL | 12 34 19.535 | −35 22 46.48 | 10.46 | 11.63 | 162.81 | 44.30 | 1999.258 | |
| CBL | 12 35 15.748 | −09 10 57.84 | 10.91 | 10.95 | 142.98 | 28.20 | 1999.089 | |

## New Common Proper-Motion Pairs from the PPMX Catalog

| Discovery Designation | RA DEC | Mags | | PA | SEP | DATE | NOTES |
|---|---|---|---|---|---|---|---|
| CBL | 12 35 43.067 −03 00 58.10 | 8.49 | 9.79 | 280.18 | 60.70 | 1999.149 | (15) |
| CBL | 12 36 16.407 −79 31 34.45 | 10.95 | 11.34 | 254.57 | 14.99 | 2000.102 | |
| CBL | 13 17 35.429 −11 57 01.34 | 11.04 | 12.70 | 344.71 | 24.29 | 1999.138 | |
| CBL | 13 53 54.390 −07 45 44.87 | 11.30 | 11.84 | 216.02 | 18.76 | 1999.163 | |
| CBL | 14 08 01.294 −13 16 09.22 | 11.84 | 12.31 | 180.53 | 13.24 | 1999.299 | |
| CBL | 14 12 31.776 −30 06 17.68 | 11.48 | 11.77 | 216.48 | 31.32 | 1999.262 | |
| CBL | 14 35 32.025 −35 26 39.17 | 11.44 | 11.72 | 211.37 | 41.19 | 2000.310 | |
| CBL | 14 37 23.205 −66 50 27.83 | 9.95 | 10.47 | 305.18 | 27.02 | 2000.258 | (16) |
| CBL | 14 53 52.152 −63 53 53.17 | 9.28 | 11.02 | 83.26 | 41.56 | 2000.223 | (17) |
| CBL | 14 55 28.254 −56 48 55.15 | 9.85 | 11.00 | 265.59 | 26.29 | 2000.146 | (18) |
| CBL | 14 58 31.045 −27 24 06.18 | 10.79 | 11.48 | 91.00 | 15.09 | 1998.486 | |
| CBL | 15 04 08.091 −26 23 26.83 | 10.97 | 11.56 | 322.81 | 26.24 | 1998.486 | |
| CBL | 15 07 12.834 −41 41 31.12 | 9.36 | 9.46 | 286.51 | 8.64 | 1999.374 | (19) |
| CBL | 15 13 59.433 −58 37 15.68 | 9.88 | 10.30 | 319.51 | 45.32 | 1999.431 | (20) |
| CBL | 16 31 42.851 +70 55 59.84 | 8.22 | 11.45 | 312.31 | 39.95 | 1999.398 | (21), (22) |
| CBL | 16 37 35.305 +69 19 17.25 | 9.07 | 10.76 | 124.95 | 99.40 | 1999.399 | |
| CBL | 17 01 49.674 +14 42 27.70 | 10.37 | 10.63 | 12.39 | 19.29 | 1999.158 | |
| CBL | 17 40 21.442 +05 43 37.44 | 11.19 | 11.98 | 342.16 | 34.69 | 2000.404 | |
| CBL | 17 54 51.203 +28 51 38.93 | 11.27 | 12.36 | 244.01 | 41.01 | 2000.204 | |
| CBL | 17 56 59.673 −46 06 31.92 | 10.91 | 10.94 | 207.38 | 11.21 | 1999.551 | |
| CBL | 17 59 53.975 −45 17 20.72 | 10.94 | 11.67 | 192.86 | 14.26 | 1999.551 | |
| CBL | 18 07 28.944 +00 29 27.19 | 11.23 | 11.54 | 355.05 | 49.82 | 1999.548 | |
| CBL | 18 13 05.162 +18 40 45.60 | 8.37 | 10.50 | 260.54 | 34.30 | 2000.209 | (23) |
| CBL | 18 21 26.185 −15 22 18.08 | 9.72 | 11.08 | 26.29 | 18.82 | 1999.333 | |
| CBL | 18 22 44.038 −40 44 59.30 | 10.32 | 10.73 | 159.99 | 12.69 | 2000.427 | |
| CBL | 18 27 24.762 +21 51 53.42 | 10.22 | 11.99 | 159.94 | 26.67 | 2000.242 | |
| CBL | 18 38 36.531 +49 00 42.13 | 9.42 | 10.72 | 260.52 | 47.47 | 1998.478 | (24) |
| CBL | 18 41 25.436 −44 32 30.63 | 10.78 | 11.14 | 53.29 | 36.30 | 2000.474 | |
| CBL | 18 45 04.604 −23 15 07.22 | 8.45 | 11.24 | 93.96 | 25.15 | 1998.478 | (25) |
| CBL | 18 54 43.138 −50 07 46.85 | 9.18 | 11.72 | 280.23 | 27.51 | 1999.726 | (26) |
| CBL | 19 01 33.281 −24 08 28.08 | 9.20 | 11.43 | 83.49 | 31.00 | 1999.262 | (27) |

*Table concludes on next page.*

## New Common Proper-Motion Pairs from the PPMX Catalog

| Discovery Designation | RA DEC | | Mags | | PA | SEP | DATE | NOTES |
|---|---|---|---|---|---|---|---|---|
| CBL | 19 07 06.084 | -14 04 09.97 | 8.64 | 10.38 | 317.61 | 20.69 | 2000.247 | (28) |
| CBL | 20 06 03.948 | -41 37 36.45 | 11.49 | 11.68 | 8.99 | 23.71 | 1999.505 | |
| CBL | 20 33 53.250 | -27 10 17.31 | 9.35 | 11.97 | 39.94 | 52.00 | 2000.561 | (29) |
| CBL | 20 36 05.730 | -67 05 22.53 | 10.50 | 11.32 | 107.53 | 24.53 | 2000.542 | |
| CBL | 20 46 51.514 | -49 28 39.62 | 10.71 | 11.39 | 197.77 | 27.90 | 1999.710 | |
| CBL | 21 10 18.359 | -13 04 05.84 | 11.13 | 11.15 | 246.88 | 16.80 | 1999.483 | |
| CBL | 21 16 13.422 | -40 40 51.94 | 11.30 | 12.07 | 151.91 | 31.23 | 1999.691 | |
| CBL | 21 29 36.229 | -44 13 50.10 | 10.36 | 10.71 | 261.71 | 90.54 | 1999.633 | |
| CBL | 21 36 58.092 | -35 53 03.02 | 7.35 | 8.60 | 265.34 | 78.45 | 2000.562 | (30) |
| CBL | 21 54 22.594 | -44 09 46.37 | 11.10 | 11.90 | 73.80 | 17.96 | 1999.718 | |
| CBL | 22 08 27.535 | -57 06 52.51 | 11.08 | 11.45 | 327.83 | 26.33 | 2000.543 | |
| CBL | 22 09 42.632 | -33 45 15.41 | 9.28 | 10.25 | 278.32 | 49.90 | 1999.560 | (31) |
| CBL | 22 32 09.402 | -13 35 51.81 | 7.72 | 9.71 | 93.97 | 41.94 | 1998.481 | (32) |
| CBL | 22 33 45.700 | +61 45 26.85 | 9.94 | 10.90 | 222.22 | 38.29 | 1999.746 | |
| CBL | 22 41 49.606 | +59 47 35.64 | 9.03 | 11.02 | 104.02 | 47.92 | 1999.741 | (33) |
| CBL | 22 47 55.497 | +03 36 07.26 | 11.28 | 11.35 | 154.24 | 23.20 | 2000.608 | |
| CBL | 22 53 55.679 | -37 09 40.50 | 10.22 | 10.59 | 313.30 | 54.44 | 1999.734 | |
| CBL | 22 54 19.523 | +30 22 18.31 | 10.46 | 11.40 | 233.61 | 14.46 | 1998.473 | |
| CBL | 23 28 08.467 | -02 26 53.36 | 8.05 | 8.43 | 226.37 | 57.33 | 1998.730 | (34), (35) |
| CBL | 23 37 40.086 | +00 46 36.37 | 10.63 | 12.3 | 156.89 | 34.18 | 2000.658 | |

Table Notes:

1. Parallax primary: 8.29 ±1.06 (Hipparcos). Spectral Type primary: F7V (Hipparcos), spectral type secondary: G (Tycho-2 Spectral Type Catalog).

2. Parallax primary: 3.57 ±0.91 (Hipparcos). Spectral Type primary: G8/K0IV (Hipparcos) .

3. Spectral Type primary: G5/8 (Tycho-2 Spectral Type Catalog).

4. Parallax primary: 12.1 ±1 (Hipparcos). Spectral Type primary: G5/G6 (Hipparcos) .

5. Parallax primary: 10.1 ±0.72 (Hipparcos). Spectral Type primary: G8/K0 (Hipparcos) .

6. Parallax primary: 7.84 ±0.71 (Hipparcos). Spectral Type primary: F0 V (Tycho-2 Spectral Type Catalog).

7. Parallax primary: 7.75 ±1.4 (Hipparcos). Spectral Type primary: K0 (Hipparcos).

8. Parallax primary: 8.94 ±0.89 (Hipparcos). Spectral Type primary: F6/F7V (Hipparcos) .

9. Spectral Type primary: G6 IV (Tycho-2 Spectral Type Catalog).

10. Spectral Type primary: K5 (Tycho-2 Spectral Type Catalog).

11. Spectral Type primary: F3/5 V (Tycho-2 Spectral Type Catalog).

12. Parallax primary: 9.94 ±0.94 (Hipparcos). Spectral Type primary: F2 (Hipparcos).

13. The secondary is TDS8273.

14. Parallax primary: 3.49 ±1.25 (Hipparcos). Spectral Type primary: K1III (Hipparcos) .

15. Parallax primary: 13.2 ±1.11 (Hipparcos).

## New Common Proper-Motion Pairs from the PPMX Catalog

*(Continued from page 211)*

Spectral Type primary: F5 (Hipparcos), spectral type secondary: K2 (Tycho-2 Spectral Type Catalog).

16. Spectral Type primary: F7/G0 (Tycho-2 Spectral Type Catalog).

17. Spectral Type primary: F2/5 III/IV (Tycho-2 Spectral Type Catalog).

18. Spectral Type primary: F8/G0 V (Tycho-2 Spectral Type Catalog).

19. Spectral Type primary: F3/5 V (Tycho-2 Spectral Type Catalog).

20. Spectral Type primary: F6/8 V (Tycho-2 Spectral Type Catalog).

21. The secondary is TDS 819.

22. Parallax primary: 7.32 ±0.61 (Hipparcos). Spectral Type primary: F5 (Hipparcos), spectral type secondary: F8 (Tycho-2 Spectral Type Catalog).

23. Spectral Type primary: G5 (Tycho-2 Spectral Type Catalog).

24. Spectral Type primary: F8 (Tycho-2 Spectral Type Catalog).

25. Parallax primary: 14.4±1.16 (Hipparcos). Spectral Type primary: G3/G5V (Hipparcos).

26. Spectral Type primary: G0 (Tycho-2 Spectral Type Catalog).

27. Spectral Type primary: G8 V (Tycho-2 Spectral Type Catalog).

28. Spectral Type primary: G0 (Tycho-2 Spectral Type Catalog).

29. Parallax primary: 15.2±1.35 (Hipparcos). Spectral Type primary: G8V (Hipparcos).

30. Parallax primary: 14.95 ±0.93(Hipparcos), parallax secondary: 15.09 ±1.21 (Hipparcos). Spectral Type primary: F3V (Hipparcos), secondary: G5V (Tycho-2 Spectral Type Catalog)..

31. Spectral Type primary: G0 V (Tycho-2 Spectral Type Catalog).

32. Parallax primary: 9.78±1.1 (Hipparcos). Spectral Type primary: G0V (Hipparcos).

33. Spectral Type primary: G0  (Tycho-2 Spectral Type Catalog).

34. Secondary is RST4724.

35. Parallax primary: 8.84 ±1.09 (Hipparcos). Spectral Type primary: F2 (Hipparcos), spectral type secondary: F5 V (Tycho-2 Spectral Type Catalog).

**Table 3**: Proper Motion of Each Component (mas/yr)

| RA  DEC | $\mu_1$ | $\mu_2$ | $\sigma_1$ | $\sigma_2$ |
|---|---|---|---|---|
| 00 32 19.099 −21 50 33.69 | (58.0, −2.6) | (63.8, 0.5) | (2.8, 3.1) | (2.8, 3.1) |
| 00 55 33.213 −43 16 11.73 | (14.3, −87.7) | (7.3, −89.7) | (2.2, 2.2) | (2.2, 2.2) |
| 01 00 52.512 −18 56 57.08 | (50.5, −7.8) | (53.2, −11.3) | (2.3, 2.3) | (2.3, 2.3) |
| 01 52 27.571 −49 31 25.56 | (31.2, −51.8) | (29.6, −50.5) | (1.5, 1.4) | (1.5, 1.4) |
| 02 04 18.761 −70 59 40.88 | (119.0, −29.9) | (124.9, −33.4) | (1.7, 2.0) | (1.9, 1.9) |
| 02 05 27.334 +38 20 57.02 | (33.1, −70.5) | (23.5, −83.2) | (2.4, 2.3) | (10.1, 10.6) |
| 03 10 41.550 −20 06 41.54 | (60.9, 3.4) | (65.7, 4.4) | (1.8, 1.8) | (2.4, 2.5) |
| 03 24 54.681 −43 12 55.77 | (3.0, 53.4) | (3.7, 58.6) | (1.8, 1.8) | (2.1, 2.1) |
| 03 46 09.569 −41 12 22.33 | (66.6, 25.0) | (71.9, 24.9) | (0.7, 0.8) | (2.1, 2.1) |
| 04 23 48.192 −76 43 09.45 | (36.4, −43.3) | (38.4, −43.5) | (1.4, 1.4) | (1.4, 1.4) |
| 05 01 36.173 −44 49 49.05 | (8.3, −65.8) | (5.2, −68.8) | (1.6, 1.6) | (2.1, 2.0) |
| 05 23 39.978 −38 18 48.09 | (29.5, 41.5) | (34.1, 37.2) | (2.3, 2.2) | (2.3, 2.3) |

*Table continues on next page.*

## New Common Proper-Motion Pairs from the PPMX Catalog

**Table 3** (continued): Proper Motion of Each Component (mas/yr)

| RA DEC | μ₁ | μ₂ | σ₁ | σ₂ |
|---|---|---|---|---|
| 05 57 24.758 −40 23 51.78 | (−1.3, 62.8) | (−3.0, 60.5) | (1.5, 1.5) | (2.3, 2.3) |
| 06 15 01.664 −63 20 38.82 | (−2.4, 84.8) | (−7.4, 83.7) | (3.8, 3.8) | (3.9, 3.9) |
| 06 38 17.669 +18 28 24.58 | (−78.3, −21.0) | (−73.4, −20.9) | (1.6, 1.6) | (1.6, 1.6) |
| 07 03 08.446 −73 50 13.91 | (−7.6, 64.2) | (−5.1, 57.0) | (6.0, 6.3) | (1.9, 1.9) |
| 07 08 55.244 −11 23 29.58 | (−5.5, −74.1) | (−2.4, −73.0) | (1.6, 1.7) | (2.0, 2.1) |
| 09 33 19.911 −07 11 24.75 | (−59.3, −26.5) | (−59.9, −23.8) | (1.5, 1.7) | (2.2, 2.2) |
| 09 51 08.004 −18 39 31.43 | (−97.4, 26.1) | (−101.1, 21.8) | (1.7, 1.6) | (3.0, 3.0) |
| 10 14 38.104 −13 33 29.10 | (−50.9, −17.8) | (−52.5, −21.1) | (1.3, 1.3) | (1.6, 1.6) |
| 10 15 08.028 −65 26 11.04 | (−62.0, 13.9) | (−60.0, 15.2) | (2.2, 2.2) | (2.2, 2.2) |
| 10 16 15.352 −17 11 13.12 | (−94.0, −27.8) | (−94.1, −35.4) | (2.6, 2.7) | (2.5, 2.5) |
| 10 32 03.297 −30 28 05.49 | (28.3, −42.4) | (38.2, −35.9) | (4.1, 4.1) | (3.4, 3.4) |
| 10 47 27.741 −11 54 08.72 | (21.8, −63.6) | (20.0, −64.3) | (1.5, 1.5) | (1.9, 2.0) |
| 11 29 03.259 −38 17 05.77 | (−9.3, −70.4) | (−12.1, −68.8) | (2.4, 2.5) | (2.9, 2.9) |
| 11 35 52.047 −40 40 36.43 | (−84.5, 8.0) | (−78.7, 10.5) | (2.8, 2.8) | (2.9, 2.9) |
| 11 53 22.088 −67 07 05.56 | (−9.1, −55.5) | (−5.9, −56.8) | (2.3, 2.3) | (2.4, 2.4) |
| 12 05 25.156 +17 17 21.69 | (−62.8, −53.3) | (−63.9, −51.8) | (1.1, 1.2) | (1.2, 1.3) |
| 12 13 30.463 −48 47 46.64 | (37.1, −54.7) | (33.7, −53.0) | (1.5, 1.5) | (2.0, 2.0) |
| 12 34 19.535 −35 22 46.48 | (−60.9, 8.7) | (−63.7, 3.6) | (2.5, 2.5) | (2.8, 2.8) |
| 12 35 15.748 −09 10 57.84 | (−49.4, −13.5) | (−48.5, −13.9) | (1.8, 1.9) | (1.9, 1.9) |
| 12 35 43.067 −03 00 58.10 | (−76.3, 13.9) | (−76.1, 14.3) | (0.7, 0.6) | (1.5, 1.5) |
| 12 36 16.407 −79 31 34.45 | (−58.2, −11.7) | (−60.5, −5.6) | (1.6, 1.6) | (2.0, 2.0) |
| 13 17 35.429 −11 57 01.34 | (−85.6, 35.8) | (−82.6, 36.2) | (1.7, 1.7) | (2.1, 2.1) |
| 13 53 54.390 −07 45 44.87 | (−48.1, −26.1) | (−45.8, −28.9) | (2.0, 2.0) | (2.7, 2.7) |
| 14 08 01.294 −13 16 09.22 | (−68.9, 0.9) | (−55.9, −15.5) | (9.7, 12.0) | (2.1, 2.1) |
| 14 12 31.776 −30 06 17.68 | (−72.7, −15.6) | (−67.0, −23.3) | (2.7, 2.7) | (2.7, 2.7) |
| 14 35 32.025 −35 26 39.17 | (−41.8, −28.4) | (−43.9, −25.8) | (2.3, 2.4) | (2.4, 2.4) |
| 14 37 23.205 −66 50 27.83 | (−56.1, −14.1) | (−55.0, −16.2) | (1.9, 1.9) | (1.8, 1.8) |
| 14 53 52.152 −63 53 53.17 | (86.2, −26.2) | (85.7, −25.7) | (2.7, 2.7) | (3.4, 3.2) |
| 14 55 28.254 −56 48 55.15 | (−46.1, −37.2) | (−51.6, −41.0) | (2.1, 2.1) | (2.2, 2.2) |
| 14 58 31.045 −27 24 06.18 | (13.8, −104.4) | (8.9, −104.8) | (3.7, 3.5) | (3.5, 3.4) |
| 15 04 08.091 −26 23 26.83 | (56.8, −8.6) | (52.8, −4.9) | (3.5, 3.2) | (2.7, 2.5) |

*Table continues on next page.*

## New Common Proper-Motion Pairs from the PPMX Catalog

**Table 3** (continued): Proper Motion of Each Component (mas/yr)

| RA DEC | μ₁ | μ₂ | σ₁ | σ₂ |
|---|---|---|---|---|
| 15 07 12.834 −41 41 31.12 | (−44.4, −25.3) | (−46.0, −28.6) | (1.8, 1.8) | (2.4, 2.4) |
| 15 13 59.433 −58 37 15.68 | (−61.8, −45.9) | (−62.2, −42.9) | (1.6, 1.6) | (2.4, 2.4) |
| 16 31 42.851 +70 55 59.84 | (−65.0, 22.5) | (−63.5, 23.4) | (1.3, 1.4) | (1.9, 1.9) |
| 16 37 35.305 +69 19 17.25 | (−82.0, 103.8) | (−82.4, 105.4) | (1.3, 1.4) | (1.8, 1.8) |
| 17 01 49.674 +14 42 27.70 | (−17.9, −47.2) | (−19.0, −47.1) | (1.6, 1.7) | (0.9, 0.9) |
| 17 40 21.442 +05 43 37.44 | (−24.4, −74.9) | (−25.2, −68.4) | (2.9, 2.9) | (2.9, 2.9) |
| 17 54 51.203 +28 51 38.93 | (−20.1, −47.0) | (−23.5, −45.9) | (1.7, 1.7) | (1.7, 1.7) |
| 17 56 59.673 −46 06 31.92 | (70.6, −20.9) | (72.6, −19.3) | (1.8, 1.8) | (1.8, 1.8) |
| 17 59 53.975 −45 17 20.72 | (5.9, −55.5) | (6.4, −53.5) | (2.1, 2.2) | (2.2, 2.3) |
| 18 07 28.944 +00 29 27.19 | (−18.2, −48.3) | (−18.4, −49.4) | (1.8, 1.8) | (1.8, 1.8) |
| 18 13 05.162 +18 40 45.60 | (5.4, −50.6) | (5.1, −52.0) | (1.1, 1.1) | (1.6, 1.6) |
| 18 21 26.185 −15 22 18.08 | (12.8, −82.3) | (12.2, −82.8) | (1.7, 1.8) | (1.8, 1.8) |
| 18 22 44.038 −40 44 59.30 | (−32.4, −38.8) | (−31.2, −40.2) | (2.2, 2.2) | (2.3, 2.4) |
| 18 27 24.762 +21 51 53.42 | (−16.4, 53.7) | (−18.0, 50.4) | (1.5, 1.5) | (2.2, 2.2) |
| 18 38 36.531 +49 00 42.13 | (69.8, 43.1) | (65.8, 42.7) | (1.4, 1.4) | (1.1, 1.1) |
| 18 41 25.436 −44 32 30.63 | (16.7, −51.9) | (14.9, −56.7) | (2.2, 2.2) | (2.2, 2.2) |
| 18 45 04.604 −23 15 07.22 | (17.7, −119.5) | (21.4, −126.1) | (1.7, 1.6) | (2.8, 2.8) |
| 18 54 43.138 −50 07 46.85 | (−2.6, −76.9) | (−1.7, −75.8) | (1.5, 1.5) | (2.2, 2.2) |
| 19 01 33.281 −24 08 28.08 | (−25.6, −83.0) | (−30.1, −82.5) | (1.6, 1.5) | (2.5, 2.6) |
| 19 07 06.084 −14 04 09.97 | (94.3, 27.6) | (91.4, 28.0) | (1.6, 1.6) | (2.2, 2.1) |
| 20 06 03.948 −41 37 36.45 | (59.1, −53.6) | (58.6, −57.6) | (2.1, 2.1) | (2.1, 2.1) |
| 20 33 53.250 −27 10 17.31 | (68.4, −82.3) | (71.2, −89.2) | (1.6, 1.6) | (3.3, 3.3) |
| 20 36 05.730 −67 05 22.53 | (−29.0, −63.8) | (−28.6, −67.9) | (1.7, 1.7) | (1.8, 1.8) |
| 20 46 51.514 −49 28 39.62 | (71.9, −2.3) | (74.1, −1.0) | (2.1, 2.2) | (2.2, 2.2) |
| 21 10 18.359 −13 04 05.84 | (73.1, −41.2) | (65.6, −42.8) | (5.2, 6.4) | (2.1, 2.1) |
| 21 16 13.422 −40 40 51.94 | (45.3, −27.1) | (46.0, −27.0) | (1.8, 1.8) | (1.8, 1.8) |
| 21 29 36.229 −44 13 50.10 | (86.8, −53.0) | (86.4, −49.2) | (2.0, 2.0) | (2.0, 2.1) |
| 21 36 58.092 −35 53 03.02 | (90.9, −10.1) | (90.7, −9.2) | (0.8, 0.5) | (1.4, 1.4) |
| 21 54 22.594 −44 09 46.37 | (13.7, −89.1) | (18.8, −92.5) | (2.1, 2.2) | (3.1, 3.1) |
| 22 08 27.535 −57 06 52.51 | (10.6, −73.0) | (11.0, −68.4) | (2.3, 2.3) | (2.4, 2.4) |
| 22 09 42.632 −33 45 15.41 | (49.3, −84.1) | (47.8, −82.7) | (1.5, 1.4) | (2.4, 2.4) |

*Table concludes on next page.*

## New Common Proper-Motion Pairs from the PPMX Catalog

**Table 3** (continued): Proper Motion of Each Component (mas/yr)

| RA DEC | $\mu_1$ | $\mu_2$ | $\sigma_1$ | $\sigma_2$ |
|---|---|---|---|---|
| 22 32 09.402 −13 35 51.81 | (−19.4, −65.9) | (−20.1, −62.8) | (1.6, 1.6) | (1.5, 1.5) |
| 22 33 45.700 +61 45 26.85 | (−18.6, −61.7) | (−15.8, −58.4) | (2.3, 2.1) | (2.0, 2.0) |
| 22 41 49.606 +59 47 35.64 | (16.9, −79.0) | (15.4, −77.5) | (1.2, 1.3) | (1.6, 1.6) |
| 22 47 55.497 +03 36 07.26 | (67.9, 3.3) | (69.6, 3.3) | (1.9, 1.9) | (1.6, 1.6) |
| 22 53 55.679 −37 09 40.50 | (108.1, 5.4) | (106.4, −0.8) | (1.4, 1.6) | (3.1, 3.1) |
| 22 54 19.523 +30 22 18.31 | (142.7, −4.6) | (142.0, −9.8) | (1.5, 1.5) | (1.6, 1.6) |
| 23 28 08.467 −02 26 53.36 | (69.8, 10.4) | (70.5, 8.4) | (1.2, 1.3) | (1.3, 1.4) |
| 23 37 40.086 +00 46 36.37 | (−30.0, −69.7) | (−29.8, −85.9) | (1.7, 1.8) | (10.7, 10.7) |

advantages of this approach are:

- The basic core operations are easier to test and debug than the usual complex operations required by the data mining process.

- Assuming that the basic operations have been thoroughly checked, the whole process becomes less prone to errors. This holds true because the different filters are now written in a higher abstraction level, instead of directly in SQL. This makes them more understandable and easier to test and modify.

- The same user-scripts can be employed in different projects, facilitating the reusability of the code. The language allows defining parameterized functions with this purpose.

- This possibility of easily modifying the code is very useful for proving variations of the same filter and designing new ones. In our case it has been crucial for comparing the two versions of the first Halbwachs' condition examined above.

Currently, we are improving the application in order to make it publicly available in the near future. Regarding our test-case, the PPMX catalog, we would like to point out two main conclusions:

- The catalog was chosen to check the software, assuming that all the interesting pairs had been already extracted. Indeed most of them were in the WDS, but still there were a few possible interesting pairs to find.

- In all the projects examined up to now many of the pairs detected by data mining did not exist in the plates. Instead they corresponded to erroneous data obtained while processing the images. However, in this catalog no false pairs were found, attesting to the quality of its data.

As usual, it is important to remark that we don't claim that the CPMPs found are true binaries. The data mining process only suggests that these pairs might deserve more measurements and a deeper study.

## Acknowledgements

This research makes use of the ALADIN Interactive Sky Atlas and of the VizieR database of astronomical catalogs, all maintained at the Centre de Données Astronomiques, Strasbourg, France, and of the data products from the Two Micron All Sky Survey, which is a joint project of the University of Massachusetts and the Infrared Processing and Analysis Center/California Institute of Technology, funded by the National Aeronautics and Space Administration and the National Science Foundation. This work has been partially supported by the Spanish projects TIN2008-06622-C03-01, S-0505/TIC/0407, S2009TIC-1465 and UCM-BSCH-GR58/08-910502.

# New Common Proper-Motion Pairs from the PPMX Catalog

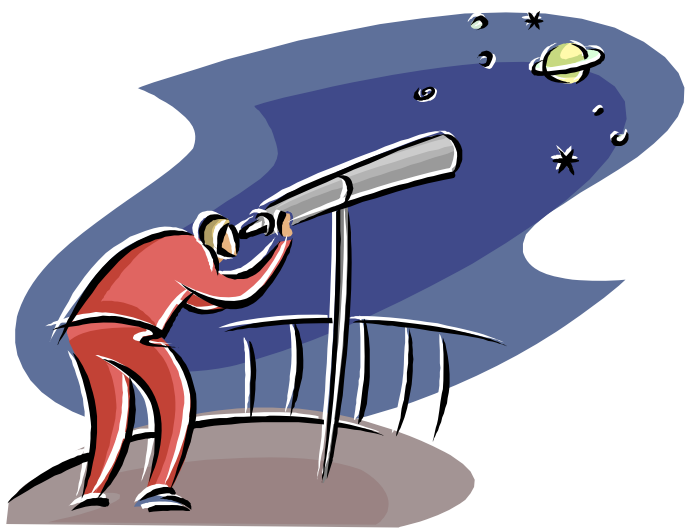